\documentclass[twocolumn,twoside,10pt]{article}
\usepackage{graphicx,amsmath,amssymb,amsthm}

\makeatletter \oddsidemargin-.25in \evensidemargin-.25in
\makeatother \topmargin-0.75in \textwidth7in \textheight9.5in

\newtheorem{thm}{thm}[section]

\newtheorem{lem}{Lemma}[section]
\newtheorem{prop}{Proposition}[section]
\theoremstyle{definition}

\newtheorem{rem}{Remark}[section]
\newtheorem{ex}{Example}[section]

\def\Proof{\medskip\noindent {\it Proof --- \ }}
\def\cqfd{\hfill $\clubsuit$ \bigskip}

\def\Q{\mathbb{Q}}

\def\card{\mathsf{card}}

\def\ra{\rightarrow}

\def\pointir{\unskip . --- \ignorespaces}

\def\cqfd{\hfill $\Box$ \bigskip}
\def\Box{\fbox{\hspace{0.5mm}}}

\def\si{\sigma}
\def\om{\omega}
\def\ep{\epsilon}

\begin{document}

\date{}

\title{\begin{flushleft}
\noindent {\small {\it Journal of Nonlinear Systems and Applications
()
\\
 Copyright $\copyright$ 2009 Watam Press \hfill http://www.watam.org/JNSA/\\[5.0mm]}}
\end{flushleft}\Large\bf \uppercase{Statistics on Graphs, Exponential Formula and Combinatorial Physics} }
\author{Laurent~Poinsot, G\'erard~H.~E.~Duchamp, Silvia~Goodenough and Karol~A.~Penson
\thanks{L. Poinsot, G. H. E. Duchamp and S. Goodenough are affiliated to Laboratoire d'Informatique Paris Nord, Universit\'e Paris-Nord 13, CNRS UMR 7030,  99 av. J.-B. Cl\'ement, F 93430 Villetaneuse, France (emails: $\{$ghed,laurent.poinsot$\}$@lipn-univ.paris13.fr, goodenou@iutv.univ-paris13.fr).}
\thanks{K. A. Penson is affiliated to Laboratoire de Physique Th\'eorique de la Mati\`ere Condens\'ee, Universit\'e Pierre et Marie Curie, CNRS UMR 7600, Tour 24 - 2e \'et., 4 pl. Jussieu, F 75252 Paris cedex 05, France (email: penson@lptl.jussieu.fr).}
\thanks{Manuscript received October 05, 2009. This work was supported by the French Ministry of Science and Higher Education under Grant ANR PhysComb.}}
 \maketitle


{\footnotesize \noindent {\bf Abstract.}
The concern of this paper is a famous combinatorial formula known under the name ``exponential formula". It occurs quite naturally in many contexts (physics, mathematics, computer science). Roughly speaking, it expresses that the exponential generating function of a whole structure is equal to the exponential of those of connected substructures. Keeping this descriptive statement as a guideline, we develop a general framework to handle many different situations in which the exponential formula can be applied. }


\vskip.2in



\section{Introduction}\label{expf}

Applying the exponential paradigm one can feel sometimes incomfortable wondering whether ``one has the right'' to do 
so (as for coloured structures, for example). The following paper is aimed at giving a rather large framework where this 
formula holds. 

Exponential formula can be traced back to works by Touchard and Ridell \& Uhlenbeck \cite{To,RU}. 
For an other exposition, see for example \cite{DM97,Fl,Jo,St}.
 
We are interested to compute various examples of EGF for combinatorial objects having 
(a finite set of) nodes (i.e. their set-theoretical support) so we use as central concept the mapping $\sigma$ which 
associates to every structure, its set of (labels of its) nodes.\\
We need to draw what could be called ``square-free decomposable objects'' (SFD). This version is suited to our needs 
for the ``exponential formula'' and it is sufficiently general to contain, as a particular case, the case of multivariate series.

\section{Partial semigroups}

Let us call partial semigroup a semigroup with a partially defined associative law (see for instance \cite{Eil74} for usual semigroups and \cite{Bru58,LE97,Seg73} for more details on structures with a partially defined binary operation). More precisely, 
a \emph{partial semigroup} is a pair $(S,*)$ where $S$ is a set and $*$ is a (partially defined) function $S\times S\rightarrow S$ such that the two (again partially defined) functions $S\times S\times S \rightarrow S$
\begin{equation}
(x,y,z)\mapsto (x*y)*z\ \mbox{and}\ (x,y,z)\mapsto x*(y*z)
\end{equation} 
coincide (same domain and values). Using this requirement one can see that the values of the (partially defined) functions $S^n \rightarrow S$
\begin{equation}
(x_1,\cdots,x_n)\mapsto E_T(x_1,\cdots,x_n)
\end{equation}
obtained by evaluating the expression formed by labelling by $x_i$ (from left to right) the $i$th leaf of a binary tree $T$ with $n$ nodes and by $*$ its internal nodes, is independant of $T$. We will denote $x_1*\cdots*x_n$ their common value.  In this paper we restrict our attention to \emph{commutative} semigroups. By this we mean that the value $x_1*\cdots*x_n$ does not depend on the relative order of the $x_i$. A nonempty partial semigroup $(S,*)$ has a \emph{(two-sided and total) unit} $\epsilon \in S$ if, and only if, for every $\om \in S$, $\om*\epsilon=\om=\epsilon*\om$. Using associativity of $*$, it can be easily checked that if $S$ has a unit, then it is unique.  
\begin{ex}
Let $F$ be a set of sets (resp. which contains $\emptyset$ as an element) and which is closed under the disjoint sum $\sqcup$, \emph{i.e.}, if $A,B\in F$ such that $A\cap B=\emptyset$, then $A \cup B (=A\sqcup B)\in F$. Then $(F,\sqcup)$ is a partial semigroup (resp. partial semigroup with unit). 
\end{ex}

\section{Square-free decomposable partial semigroups}
Let $\mathsf{2}^{(\mathbb{N}^+)}$ be the set of all finite subsets of the positive integers $\mathbb{N}^+$ and $(S,\oplus)$ be a partial semigroup with unit (here denoted $\epsilon$) equipped with a mapping $\si:S \rightarrow \mathsf{2}^{(\mathbb{N}^+)}$, called the \emph{(set-theoretic) support mapping}. Let $D$ be the domain of $\oplus$. The triple $(S,\oplus,\si)$ is called \emph{square-free decomposable} (SFD) if, and only if, it fulfills the two following conditions. 
\begin{itemize}
\item \emph{Direct sum} (DS): 
\begin{enumerate}
\item $\sigma(\om)=\emptyset$ iff $\om=\epsilon$;
\item $D=\{(\om_1,\om_2)\in S^2 : \sigma(\om_1)\cap \sigma(\om_2)=\emptyset\}$;
\item For all $\om_1,\om_2\in S$, if $(\om_1,\om_2)\in D$ then $\sigma(\om_1\oplus \om_2)=\sigma(\om_1)\cup\sigma(\om_2)$.
\end{enumerate}
\item \emph{Levi's property} (LP): For every $\om_1,\om_2,\om^1,\om^2 \in S$ such that $(\om_1,\om_2),(\om^1,\om^2)\in D$ and $\om_1\oplus \om_2=\om^1\oplus \om^2$, there are $\om_i^j\in S$ for $i=1,2$, $j=1,2$ 
such that $(\om_i^1,\om_i^2),(\om_1^j,\om_2^j)\in D$, 
$\om_i=\om_i^1\oplus\om_i^2$ and $\om^j=\om^j_1\oplus \om^j_2$ for $i=1,2$ and $j=1,2$. 
\end{itemize}

\begin{rem}
The second and third conditions of (DS) imply that $\si(\om_1\oplus \om_2)=\si(\om_1)\sqcup \si(\om_2)$ whenever $(\om_1,\om_2)\in D$ (which means that $\si(\om_1)\cap\si(\om_2)=\emptyset$), where $\sqcup$ denotes the disjoint sum. 
\end{rem}
\begin{ex}
As example of this setting we have:\\
\begin{enumerate}\label{ex1}
\item\label{intsf} The positive square-free integers, $\si(n)$ being the set of primes which divide $n$, the atoms being the 
prime numbers.
\item\label{int} All the positive integeres ($S=\mathbb{N}^+$), under the usual integer multiplication, $\sigma(n)$ being the set of primes which divide $n$.
\item\label{graphs} Graphs, hypergraphs, (finitely) coloured, weighted graphs, with nodes in $\mathbb{N}^+$, $\si(G)$ being the set of nodes and $\oplus$ the 
juxtaposition (direct sum) when the set of nodes are mutually disjoint. 
\item\label{endof} The set of endofunctions $f:F\rightarrow F$ where $F$ is a finite subset of $\mathbb{N}^+$.
\item\label{ex13} The (multivariate) polynomials in $\mathbb{N}[X]$, $X=\{x_i : i\in I\}$, with $I\subseteq \mathbb{N}^+$, being a nonempty set of (commuting or not) variables, with $\si(P)=Alph(P)$ the set of indices of variables that occur in a polynomial $P$,  and $\oplus=+$.
\item\label{ex14} For a given finite or denumerable field, the set of irreducible monic polynomials is denumerable. Arrange them in a sequence $(P_n)_{n\in\mathbb{N}^+}$, then the square-free monic (for a given order on the variables) polynomials is SFD,   $\si(P):=\{n\in\mathbb{N}^+ : P_n\ \mathit{divides}\ P\}$ and $\oplus$ being the multiplication.
\item\label{ex15} Rational complex algebraic curves; $\si(V)$ being the set of monic irreducible bivariate polynomials vanishing on $V$. 
\end{enumerate}
\end{ex}

In what follows we write $\displaystyle\oplus_{i=1}^n \om_i$ instead of $\om_1\oplus\dots\oplus\om_n$ (if $n=0$, then $\oplus_{i=1}^n \om_i=\epsilon$) and  we suppose that $(S,\oplus,\sigma)$ is SFD for the two following lemmas.
\begin{lem}\label{lemme_somme_generale_et_support}
Let $\om_1,\dots,\om_n\in S$ such that $\displaystyle\oplus_{i=1}^n \om_i$ is defined. Then for every $i,j\in \{1,\dots,n\}$ such that $i\not=j$, it holds that $\sigma(\om_i)\cap\sigma(\om_j)=\emptyset$.  In particular, if none $\om_k$ is equal to $\epsilon$, then $\om_i\not=\om_j$ for every $i,j\in\{1,\dots, n\}$ such that $i\not=j$.  Moreover  $\sigma(\displaystyle\oplus_{i=1}^n \om_i)=\bigsqcup_{i=1}^n \sigma(\om_i)$. 
\end{lem}
 

\begin{lem}
Let $(\omega_i)_{i=1}^n$ be a finite family of elements of $S$ with pairwise disjoint supports.  Suppose that for $i=1,\cdots, n$, $\omega_i=\displaystyle\oplus_{k=1}^{n_i}\om_{i}^{k}$, where $(\om_i^{k})_{k=1}^{n_i}$ is a finite family of elements of $S$. Then 
$\displaystyle\oplus_{i=1}^n \omega_i=\oplus_{i=1}^n \left ( \oplus_{k=1}^{n_i}\om_{i}^{k}\right )$. 
\end{lem}
These lemmas are useful to define the sum of two or more elements of $S$ using respective sum decompositions.\\
 
Now, an \emph{atom} in a partial semigroup with unit $S$ is any object $\om\not=\ep$ which cannot be split, formally 
\begin{equation}
\om=\om_1\oplus \om_2\Longrightarrow \ep\in \{\om_1,\om_2\}\ .
\end{equation}
The set of all atoms is denoted by $\mathtt{atoms}(S)$. 
Whenever the square-free decomposable semigroup $S$ is not \emph{trivial}, \emph{i.e.}, reduced to $\{\epsilon\}$, $\mathtt{atoms}(S)$ is not empty. 
\begin{ex}
The atoms obtained from examples~\ref{ex1}:
\begin{enumerate}\label{ex2}
\item The atoms of \ref{ex1}.\ref{int} are the primes.
\item The atoms of \ref{ex1}.\ref{graphs} are connected graphs.
\item The atoms of \ref{ex1}.\ref{endof} are the endofunctions for which the domain is a singleton.
\item The atoms of \ref{ex1}.\ref{ex13} are the monomials.  
\end{enumerate}
\end{ex}

The prescriptions (DS,LP) imply that decomposition of objects into atoms always exists  and is unique. 
\begin{prop}\label{decomposition_atomique_unique}
Let $(S,\oplus,\sigma)$ be SFD. For each $\om \in S$ there is one and only one  finite set of atoms $A=\{\om_1,\cdots,\om_n\}$ such that $\om={\oplus}_{i=1}^n \om_i$. One has $A=\emptyset$ iff $\omega=\epsilon$. 
\end{prop}

\section{Exponential formula}
In this section we consider $(S,\oplus,\sigma)$ as a square-free decomposable partial semigroup with unit.\\

In the set $S$, objects are conceived to be  ``measured'' by different parameters (data in statistical language). So, to get 
a general purpose tool, we suppose that the statistics takes its values in a (unitary) ring $R$ of characteristic zero that is to say which contains $\Q$ (as, to write 
exponential generating series it is convenient to have at hand the fractions $\frac{1}{n!}$). Let then $c: S\ra R$ be the 
given statistics. For $F$ a finite set and each $X\subseteq S$, we define 
\begin{equation}
X_F:=\{\om\in X : \sigma(\om)=F\}\ .
\end{equation}
In order to write generating series, we need 
\begin{enumerate}
\item that the sums $\displaystyle c(X_F):=\sum_{\om\in X_F}c(\om)$ exist for every finite set $F$ of $\mathbb{N}^{+}$ and every $X\subseteq S$;
\item that $F\ra c(X_F)$ would depend only of the cardinality of the finite set $F$ of $\mathbb{N}^+$, for each fixed $X\subseteq S$;
\item that $c(\om_1\oplus \om_2)=c(\om_1).c(\om_2)$.
\end{enumerate}
We formalize it in \\

\noindent (LF) \emph{Local finiteness}\pointir For each finite set $F$ of $\mathbb{N}^+$, the subset $S_F$ of $S$ 
is a finite set.\\
(Eq) \emph{Equivariance}\pointir 
\begin{equation}
\card(F_1)=\card(F_2)\Longrightarrow c(\mathtt{atoms}(S)_{F_1})=c(\mathtt{atoms}(S)_{F_2})\ .
\end{equation}
(Mu) \emph{Multiplicativity}\pointir 
\begin{equation}
c(\om_1\oplus \om_2)=c(\om_1).c(\om_2)\ .
\end{equation}

\begin{rem}
a) In fact, (LF) is a property of the set $S$, while (Eq) is a property of the statistics. 
In practice, we choose $S$ which is locally finite and choose equivariant statistics for instance
$$
c(\om)=x^{\rm (number\ of\ cycles)}y^{\rm (number\ of\ fixed\ points)}
$$ 
for some variables $x,y$.\\

b) More generally, it is typical to take integer-valued partial (additive) statistics $c_1, \cdots c_i,\cdots ,c_r$ 
(for every $\omega\in S$, $c_i(\omega)\in \mathbb{N}$) and set
$c(\om)=x_1^{c_1(\omega)}x_2^{c_2(\omega)}\cdots x_r^{c_r(\omega)}$.\\
 
c) The set of example \ref{ex1}.\ref{int} is not locally finite, but other examples satisfy (LF): for instance \ref{ex1}.\ref{graphs} 
if one asks that the number of arrows and weight is finite, \ref{ex1}.\ref{intsf}.
\end{rem}

A multiplicative statistics is called \emph{proper} if $c(\epsilon)\not=0$. It is called \emph{improper} if $c(\epsilon)=0$. In this case, for every $\om\in S$, $c(\om)=0$ as 
$c(\om)=c(\om\oplus \epsilon)=c(\om)c(\epsilon)=0$. 

If $R$ is a integral domain and if $c$ is proper, then $c(\epsilon)=1$ because $c(\epsilon)=c(\epsilon\oplus\epsilon)=c(\epsilon)^2$, therefore $1=c(\epsilon)$. Note that for each $X\subseteq S$,  $c(X_{\emptyset})=\displaystyle\sum_{\om\in X_{\emptyset}}c(\om)=\left\{\begin{array}{lll}
c(\epsilon)&\mathit{if} & \epsilon\in X\\
0 & \mathit{if} & \epsilon\not\in X\end{array}\right .$. For every finite subset $X$ of $S$, we also define $c(X):=\displaystyle\sum_{\om\in X}c(\om)$, then we have in particular $c(\emptyset)=0$ (which is not the same as
 $c(S_{\emptyset})=c(\{\epsilon\})$ if $c$ is proper). The requirement (LF) implies that for every $X\subseteq S$ and every finite set $F$ of $\mathbb{N}^+$, $c(X_F)$ is defined as a sum of a finite number of terms because $X_F \subseteq S_F$, and therefore $X_F$ is finite. \\

Now, we are in position to state the exponential formula as it will be used throughout the paper. Let us recall the usual \emph{exponential formula} for formal power series in $R[[z]]$ (see~\cite{La,St} for more details on formal power series). Let $f(z)=\displaystyle\sum_{n\geq 1}f_n z^n$. Then we have 
\begin{equation}
e^f=\displaystyle\sum_{n\geq 0}a_n \frac{z^n}{n!}
\end{equation} 
where 
\begin{equation}
a_n = \displaystyle\sum_{\pi\in \Pi_n}\prod_{p\in\pi}f_{\card(p)}
\end{equation}
with $\Pi_n$ being the set of all partitions of $[1..n]$ (in particular for $n=0$, $a_0=1$) and $e^z=\displaystyle\sum_{n\geq 0}\frac{z^n}{n!}\in R[[z]]$. \\
In what follows $[1..n]$ denotes the interval $\{j\in \mathbb{N}^+ : 1\leq j\leq n\}$, reduced to $\emptyset$ when $n=0$. Let $(S,\oplus,\si)$ be a locally finite SFD and $c$ be a multiplicative equivariant statistics. For every subset $X$ of $S$ one sets the following \emph{exponential
generating series}
\begin{equation}
\mathbf{EGF}(X; z) = \displaystyle\sum_{n= 0}^{\infty}c(X_{[1..n]})\frac{z^n}{n!}\ .
\end{equation}
\begin{thm}[exponential formula]\label{expformula}
Let $S$ be a locally finite SFD and $c$ be a multiplicative equivariant statistics. We have 
\begin{equation}
\mathbf{EGF}(S;z)=c(\epsilon)-1+e^{\mathbf{EGF}({\mathtt{atoms}(S)};z)}\ ..
\end{equation}
In particular if $c(\epsilon)=1$ (for instance if $c$ is proper and $R$ is an integral domain), 
\begin{equation}
\mathbf{EGF}(S;z)=e^{\mathbf{EGF}({\mathtt{atoms}(S)};z)}\ .
\end{equation}
\end{thm}

\Proof
Let $n=0$. Then the unique element of $S_{\emptyset}$ is $\epsilon$. Therefore $c(S_{\emptyset})=c(\epsilon)$. Now suppose that $n> 0$ and let $\omega\in S_{[1..n]}$. According to proposition~\ref{decomposition_atomique_unique}, there is a unique finite set $\{\alpha_1,\dots,\alpha_k\}\subseteq {\mathtt{atoms}(S)}$ such that $\omega=\displaystyle\oplus_{i=1}^k \alpha_i$. By 
lemma~\ref{lemme_somme_generale_et_support}, $\{\sigma(\alpha_i):1\leq i\leq k\}$ is a partition of $[1..n]$ into $k$ blocks. Therefore $\omega\in {\mathtt{atoms}(S)}_{P_1}\oplus\dots\oplus {\mathtt{atoms}(S)}_{P_k}$ where $P_i=\sigma(\alpha_i)$ for $i=1,\dots,k$. We can remark that $\alpha_1\oplus\dots\oplus\alpha_k$ is well-defined for each $(\alpha_1,\dots,\alpha_k)\in {\mathtt{atoms}(S)}_{P_1}\times\dots\times {\mathtt{atoms}(S)}_{P_k}$ since the supports are disjoint. 
Now, one has, thanks to the partitions of $[1..n]$
\begin{eqnarray}
\displaystyle S_{[1..n]}=\bigsqcup_{\pi\in \Pi_n}\bigoplus_{p\in\pi}{\mathtt{atoms}(S)}_p \\
c(S_{[1..n]})=\sum_{\pi\in\Pi_n}\prod_{p\in \pi}c({\mathtt{atoms}(S)}_{p})
\end{eqnarray}
as, for disjoint (finite) sets $F$ and $G$ of $\mathbb{N}^+$, it is easy to check that $c(X_F\oplus X_G)=c(X_F)c(X_G)$ for every $X\subseteq S$ and because the disjoint union as only a finite number of factors. 
Therefore due to equivariance of $c$ on sets of the form ${\mathtt{atoms}(S)}_{F}$, one has 
\begin{equation}
\displaystyle c(S_{[1..n]})=\sum_{\pi\in\Pi_n}\prod_{p\in \pi}c({\mathtt{atoms}(S)}_{[1..\card(p)]})\ .
\end{equation}
But $c({\mathtt{atoms}(S)}_{[1..\card(p)]})$ is the $\card(p)$th coefficient of the series $\mathbf{EGF}({\mathtt{atoms}(S)};z)$. Therefore due to the usual exponential formula, $\mathbf{EGF}(S;z)=c(\epsilon)-1+e^{\mathbf{EGF}({\mathtt{atoms}(S)};z)}$. Now if  $c(\epsilon)=1$, then we obtain $\mathbf{EGF}(S;z)=e^{\mathbf{EGF}({\mathtt{atoms}(S)};z)}$. 
\cqfd

\section{Two examples}

The examples provided here pertain to the class of labelled graphs where the ``classic" exponential formula applies, namely \emph{Burnside's Classes\footnote{The name is related to the notion of \emph{free Burnside semigroups}, namely the quotient of the free semigroup $A^{+}$, where $A$ is a finite alphabet, by the the smallest congruence that contains the relators $\om^{n+m}=\om^n$, $\om\in A^{+}$. For more details see~\cite{PdLS01}.}} $\mathit{Burn}_{a,b}$, defined, for $0\leq a < b$ two integers, as the class of graphs of numeric endofunctions $f$ such that 
\begin{equation}
f^a = f^b
\end{equation}
where $f^n$ denotes the $n$th power with respect to functional composition. Despite of its simplicity, there are still (enumerative combinatorial) open problems for this class and only $B_{1,\ell+1}$ gives rise to an elegant formula \cite{GJ,St} (see also  \cite{HS67}, for the idempotent case: $\ell=1$ and compare to exact but non-easily tractable formulas in \cite{DM97} for the general case in the symmetric semigroup, and in \cite{Krat} for their generalization to the wreath product of the symmetric semigroup and a finite group).\\

The second example: the class of finite partitions which can be (and should here) identified as graphs of equivalence relations on finite subsets $F\subseteq \mathbb{N}^+$. Call this class ``Stirling class" as the number of such graphs with support $[1..n]$ and $k$ connected components is exactly the Stirling number of the second kind $S_2(n,k)$ and, using the statistics $x^{(\mathit{number\ of\ points})}y^{(\mathit{number\ of\ connected\ components})}$, one obtains 
\begin{equation}\label{stirling-class-classique}
\displaystyle\sum_{n,k\geq 0}S_2(n,k)\frac{x^n}{n!}y^k=e^{y(e^x-1)}\ .
\end{equation}

Examples of this kind bring us to the conclusion that bivariate statistics like $Burn_{a,b}(n,k)$, 
$S_2(n,k)$ or $S_1(n,k)$ (Stirling numbers of the second and first kind) are better understood 
through the notion of one-parameter group, conversely such groups naturally arinsing in 
Combinatorial Physics lead to such statistics and new ones some of which can be interpreted 
combinatorially.

\section{Generalized Stirling numbers in Combinatorial Physics}

In Quantum Mechanics, many tools boil down to the consideration of creation and annihilation operators which will be here denoted respectively $a^{\dagger}$ and $a$. These two symbols do not commute and are subject to the unique relation 
\begin{equation}
[a,a^{\dagger}]=1\ .
\end{equation}
The complex algebra generated by these two symbols and this unique relation, the \emph{Heisenberg-Weyl algebra}, will be here denoted by $HW_{\mathbb{C}}$. The consideration of evolution (one-parameter) groups $e^{\lambda \Omega}$ where 
$\Omega=\displaystyle\sum_{\omega\in HW_{\mathbb{C}}}\alpha(\omega)\omega$ is an element of $HW_{\mathbb{C}}$, with all - but a finite number of them - the complex numbers $\alpha(\omega)$ equal to $0$, and $\omega$ a word on the alphabet $\{a,a^{\dagger}\}$ leads to the necessity of solving the \emph{Normal Ordering Problem}, \emph{i.e.}, the reduction of the powers of $\Omega$ to the form 
\begin{equation}
\Omega^n = \displaystyle\sum \beta_{i,j}(a^{\dagger})^i a^j\ .
\end{equation}
In the sequel, $\mathit{Normal}(\Omega^n)$ denotes such a sum. 
This problem can be performed with three indices in general and two in the case of homogeneous operators that is operators for which the ``excess" $e=i-j$ is constant along the monomials $(a^{\dagger})^ia^j$ of the support (for which $\beta_{i,j}\not=0$). Thus, for 
\begin{equation}
\Omega = \displaystyle\sum_{i-j=e}\beta_{i,j}(a^{\dagger})^ia^j
\end{equation}
one has, for all $n\in\mathbb{N}$, 
\begin{equation}
\mathit{Normal}(\Omega^n)=(a^{\dagger})^{ne}\displaystyle\sum_{k=0}^{\infty}S_{\Omega}(n,k)(a^{\dagger})^ka^k
\end{equation}
when $e\geq 0$, and 
\begin{equation}
\mathit{Normal}(\Omega^n)=\displaystyle\left ( \sum_{k=0}^{\infty}S_{\Omega}(n,k)(a^{\dagger})^ka^k\right )a^{n|e|}
\end{equation}
otherwise. It turns out that, when there is only one annihilation, one gets a formula of the type ($x,y$ are formal commutative variables) 
\begin{equation}
\displaystyle\sum_{n,k\geq 0}S_{\Omega}(n,k)\frac{x^n}{n!}y^k = g(x)e^{y\sum_{n\geq 1}S_{\Omega}(n,1)\frac{x^n}{n!}}
\end{equation} 
which is a generalization of formula (\ref{stirling-class-classique}). A complete study of such a procedure and the details to perform the solution of the normal ordering problem may be found in \cite{GHED}. 

\section{Conclusion}

In this paper, we have broadened \footnote{A part of our setting 
can be reformulated in the categorical context \cite{BLL,CKM}}$^{,}$
\footnote{Another direction is the $q$-exponential formula \cite{Ge,Qu}.} the domain
of application of the exponential formula, a
tool originated from statistical physics.
This broadening reveals us, together with the
essence of ``why this formula works'', a possibility
of extension to denominators other than the
factorial and, on the other hand, provides a 
link with one-parameter groups 
whose infinitesimal generators are (formal) vector
fields on the line. The general combinatorial
theory of the correspondence
(vector fields $\leftrightarrow$ bivariate statistics)
is still to be done despite the fact that we have
already a wealth of results in this direction.

\section*{Acknowledgements}
We would like to thank Christian Krattenthaler (from Wien) for fruitful discussions.\\ 
The research of this work was supported, in part, by the Agence Nationale de la Recherche (Paris, France) under Program No. ANR-08-BLAN-0243-2. We would like also to acknowledge support from ``Projet interne au LIPN 2009'' ``Polyz\^eta functions''. 

\footnotesize


\end{document}